\documentclass{aa}  
\usepackage{graphicx}
\usepackage{multirow}
\usepackage{xcolor  }
\usepackage{multicol}
\usepackage{txfonts }
\usepackage{natbib  }
\usepackage{hyperref}
\usepackage{breakurl}
\usepackage{ulem    }
\newcommand{\RA}[3]{  $#1^\mathrm{h} #2^\mathrm{m} #3^\mathrm{s}$}
\newcommand{\Dec}[3]{ $#1^\circ      #2^\prime     #3^{\prime\prime}$}

\newcommand{\Jyb}{\textrm{Jy~beam}$^{-1}$}
\newcommand{\kms}{km~s$^{-1}$}
\newcommand{\amin}{$^{\prime}$}
\newcommand{\asec}{$^{\prime\prime}$}

\newcommand{\cm}[1]{$\mathrm{cm}^{#1}$}
 \definecolor{verde}{rgb}{0.0,0.4,0.0}

\def\citeapos#1{\citeauthor{#1}'s (\citeyear{#1})}

\def\46{{G46.8--0.3}}

\begin{document}

\newcommand\tacho{\bgroup\markoverwith{\textcolor{red}{\rule[0.5ex]{2pt}{1pt}}}\ULon}

\title{Supernova remnant {\46}, a new case of interaction with molecular material}
\author{L. Supan\inst{1}
   \and G. Fischetto\inst{2}
   \and G. Castelletti\inst{1}}
\institute{Instituto de Astronom\'ia y F\'isica del Espacio (IAFE, CONICET $-$ UBA) CC 67, Suc. 28, 1428 Buenos Aires, Argentina\\
\email{lsupan@iafe.uba.ar}
\and Universidad Nacional del Sur (UNS), Av. L. N. Alem 1253, B8000CPB - Bah{\'i}a Blanca, Argentina}

\date{Received DD October 2021 / accepted 08 March 2022}
\abstract
{The Galactic supernova remnant (SNR) \object{G46.8--0.3} has been known for more than fifty years, in spite of which no specific works about this source nor on its environment have been published to date. 
To make progress on this matter, we measured new flux densities from radio surveys and combined them with previous estimates carefully collected from the literature to create an improved and fully-populated version of the integrated radio spectrum for {\46}. The resulting spectrum exhibits a featureless power-law form with an exponent $\alpha = -0.535 \pm 0.012$. The lack of a spectral turnover at the lowest radio frequencies, observable in many other SNRs, excludes the presence of abundant ionised gas either proximate to the SNR itself, or along its line of sight. 
The analysis of local changes in the radio spectral index across {\46} suggests a tendency to slightly steepen approximately at 1~GHz. Even if it is real, it does not impact on the integrated spectrum of the source. 
Deeper imaging of the radio structures of {\46} and spectral maps constructed from matched raw data are needed to provide new insights into the local spectral properties of the remnant. 
On the basis of the spectral properties of the atomic gas we placed the remnant at $8.7 \pm 1.0$~kpc and, as a bonus, we revisited the distance to the nearby \ion{H}{ii} region \object{G046.495$-$00.241} to $7.3 \pm 1.2$~kpc. From evolutionary models and our distance estimate, we conclude {\46} is a middle-aged ($\sim$1 $\times10^4$~yr) SNR. 
Furthermore, we recognised several \element[][12]{CO} and \element[][13]{CO} molecular structures in the proximity of the remnant. We used combined CO-\ion{H}{i} profiles to derive the kinematic distances to these features and characterise their physical properties. We provided compelling evidences for environmental molecular clouds physically linked to {\46} at its centre, on its eastern edge, and towards the northern and southwestern rims on the far side of the SNR shell. 
Our study of the molecular matter does not confirm that the remnant is embedded in a molecular cavity as previously suggested. {\46} shows a line-of-sight coincidence with the $\gamma$-ray source 4FGL~J1918.1+1215c detected at GeV energies by the space telescope {\it Fermi}. 
A rough analysis based on the properties of the interstellar matter close {\46} indicates that the  GeV $\gamma$-rays photons detected in direction to the SNR can be plausibly attributed to hadronic collisions and/or bremsstrahlung radiation.}

\keywords{
ISM: supernova remnants -- 
ISM: individual objects: G46.8--0.3 --
radio continuum: general -- 
gamma rays: ISM}
\titlerunning{Radio properties of the SNR~{\46} and its environs}
\authorrunning{Author et al.}
\maketitle

\section{Introduction}
\label{intro}
The distribution and density of the ambient medium structures unquestionably represent key elements shaping the evolution of supernova remnants (SNRs). Atomic and molecular constituents of the interstellar medium (ISM) have  been robustly detected in the vicinity of  remnants from both  core-collapse (CC, Type II or Ib/c SNe; see, for instance, \citealt{kuriki2018-kes79-CC}, \citealt{supan2018}, \citealt{sano2020-RXJ1713}, \citealt{sano2020-N132-CC}) and those created by the deflagration of an accreting white dwarf (two emblematic examples are the Type Ia SNR  Tycho in our Galaxy, \citealt{zhou2016-tycho-Ia-Cocavity}, \citealt{chen2017-Tycho-Ia}, and N103B in the Large Magellanic Cloud, \citealt{sano2018-N103B-Ia}). 
Reciprocally, the physical properties of the interstellar material can drastically be modified by the passage of SN long-lived shocks. But this is not the complete story because the synergy with the medium can even begin before the SN event. Indeed, several SNRs have been observed expanding into shells and cavities carved by outflow phenomena in the progenitor stars. 
Examples of this are the CC SNRs 
Kes~75 \citep{su2009-shellCO-kes75}, and 
CTB~87 \citep{liu2018-ctb87-HIcavity}, as well as the SN Ia remnants 
Kepler \citep{chiotellis2012-kepler-Ia-cavity-model} and 
RCW~86 \citep{sano2019-RCW86-Ia}. 

Observations also show that a good fraction of well-known interacting SNRs are bright in $\gamma$ rays at GeV energies \citep{tang2019}. Interpreted as a manifestation of accelerated hadrons, this result can help improving our understanding of how the interplay between environmental properties and  physical conditions at the SNR shock impacts on the production and propagation of cosmic rays. 

In this work we look in detail at the relationship between SNR~{\46} (also known as HC~30) and the interstellar matter around it. The earliest view at radio wavelengths of this remnant was presented by \citet{willis73}. 
Since then, however, no works specially devoted to the remnant or its immediate environment have been published. Some characteristics of the radio emission from {\46} are only recorded in papers dealing with a sample of SNRs (e.g. \citealt{kassim-89-list}, \citealt{dubner96}, \citealt{sun11}, \citealt{ranasinghe2018-distance}). 
Concerning the ambient medium, the only information presented to date comes from the atlas of CO-line structures carried out by \citet{sofue2021} on a large population of Galactic SNRs including {\46}. These authors pointed out that a SNR-molecular cloud scenario is feasible in direction to {\46}, as they found evidences suggesting it evolves in a molecular shell at 52~km~s$^{-1}$. Adding to the picture described above, $\gamma$-ray emission has been observed at GeV energies by the  \it Fermi \rm Gamma-ray Space Telescope  in the direction of {\46} \citep{abdollahi2020-fermi4thcatalog} but its relationship with the remnant has not been studied to date.  

This paper is organised as follows. The description of data used to analyse the radio structure of \46, as well as those corresponding to the atomic and molecular components of the interstellar matter in direction to the remnant is provided in Sect.~\ref{data}. Then, in Sect.~\ref{thor}, we examine the radio continuum spectral properties of the remnant. The analysis of the spatial distribution of the atomic material in the region of \46, along with a revised distance determination to our target source and the nearby \ion{H}{ii} region G046.495$-$00.241  are given in Sect.~\ref{HI}. This section also includes the estimation of the {\46}'s age made from standard evolutionary models. 
The spatial distribution of the molecular gas in the {\46} direction and the spectral properties of the unveiled CO structures are discussed in Sect.~\ref{CO}. 
In Sect.~\ref{gamma}, we explore possible scenarios for the GeV $\gamma$ rays detected in the region of the remnant. We close summarising our findings in Sect.~\ref{summary}.

\section{Data}
\label{data}
\subsection{Radio continuum emission}
\label{radio}
We used continuum data from The HI/OH/Recombination Line Survey of the Inner Milky Way (THOR,\footnote{\url{https://www2.mpia-hd.mpg.de/thor/Overview.html}} \citealt{beuther+16}) to trace the forward shock of {\46} at 1.4~GHz. This image represents the best view at radio wavelengths of the remnant presented to date. It includes both interferometric observations from the Karl G. Jansky Very Large Array combined with data from The VLA Galactic Plane Survey (VGPS, \citealt{stil+06}), which in turn includes short spacing data from the Effelsberg Radio Telescope. The angular resolution of the 1.4-GHz image is 25{\asec} and its rms noise level  is 0.72~m{\Jyb}. 

In order to track down possible changes in the radio spectral index as a function of frequency and position across the remnant, we constructed maps from the direct ratio between the 1.4-GHz THOR+VGPS image with the ones at 200~MHz (HPBW $2.^{\prime}9 \times 2.^{\prime}4$, rms $\sim$ 30~mJy~beam$^{-1}$) extracted from The Galactic and Extragalactic All-Sky Murchison Widefield Array Survey (GLEAM,\footnote{\url{https://www.mwatelescope.org/gleam}.} \citealt{wayth15}, \citealt{hurley19}), and at 4.8~GHz (HPBW $3.^{\prime}6 \times 3.^{\prime}4$, rms $\sim$ 9~mJy~beam$^{-1}$) from The Green Bank Northern Sky Survey (GB6, \citealt{gregory96}). 
Before combining the radio images, all of them were convolved to a resolution of 4{\amin} and also aligned and interpolated to have a pixel-by-pixel match to each other. 
Even though they were not matched in the $uv$-plane, we highlight that the resulting maps are still quite appropriate to reveal trends in the spectral index distribution across {\46}.

\subsection{Interstellar gas data and spectral analysis}
\label{ism}
Firstly, we notice that throughout this paper the velocities of the atomic and molecular line emissions are always referred to that of the local standard of rest, LSR. The mean error in our radial velocity measurements is   7.6~km~s$^{-1}$, estimated from the combined effects of streaming motions, the spectral resolution of the data, and the uncertainty in determining the central velocity. All of these contributions were added in quadrature. 

Regarding the atomic component of the ISM, it was analysed through the neutral hydrogen (\ion{H}{i}) 21~cm line emission. The data used to this purpose were extracted from VGPS\footnote{\url{http://www.ras.ucalgary.ca/VGPS/index.html}} \citep{stil+06},  which combines observations carried out with the VLA and data from the 100-m single-dish Green Bank Telescope at the NRAO. The beam size  of the observations is $\sim$60{\asec}. The spectral resolution and the typical rms noise level per channel are 1.56~{\kms} and $\sim$2~K, respectively.

To characterise the molecular environment of {\46}, we used observations in the rotational transition emission $J$ = 1-0 of the carbon monoxide isotopologues \element[ ][12]{CO} and \element[ ][13]{CO} taken from the publicly available FOREST Unbiased Galactic Plane Imaging Survey constructed with the Nobeyama 45-m telescope (FUGIN,\footnote{\url{https://nro-fugin.github.io/}} \citealt{umemoto+17}). 
The angular resolution of the data is $\sim$20{\asec}, with a sensitivity of 0.24~K for \element[ ][12]{CO} and 0.12~K for \element[ ][13]{CO}. For both CO isotopologues lines  the velocity resolution of the data is 1.3~{\kms}, with a separation between consecutive velocity channels of 0.65~{\kms}.

In our analysis we used molecular line measurements of \element[ ][12]{CO} to estimate the column density of the cold molecular hydrogen $N_\mathrm{H_2}$. We have derived  $N_\mathrm{H_2}$ values using the conversion factor recommended by \citet{bolatto2013}, $2\times10^{20}$~cm$^{-2}$~(K~km~s$^{-1}$)$^{-1}$, between $N_{\mathrm{H_{2}}}$ and the integrated \element[ ][12]{CO} $J$= 1-0 surface brightness $W_{\element[ ][12]{CO}}=\int{T_{\mathrm{B}}(v)\,dv}~\mathrm{km~s^{-1}}$, where $T_{\mathrm{B}}(v)$ is  the \element[ ][12]{CO} brightness temperature. 

By using the estimation made for $N_{\mathrm{H_{2}}}$, the mass of the molecular features (assumed mostly consisting of molecular hydrogen) discovered in the surveyed region around {\46} SNR comes from the relation $M = \mu \, m_\mathrm{H} \, d^2 \, \Omega \, N_\mathrm{H_2}$. Here $\mu=2.8$ is the mean molecular weight if a relative helium abundance of 25\% is assumed, $m_\mathrm{H}$ is the hydrogen mass, and $\Omega$ represents the solid angle along the light of sight subtended by the molecular structure placed at a distance $d$. 
In addition, the number density of each cloud is obtained through the expression $n_{\mathrm{H_{2}}} = N_\mathrm{H_{2}} / L$, where $L$ is the depth of the molecular cloud in the line of sight, which was assumed to be equal to its average size in the plane of the sky. The physical properties of the molecular features identified in direction to {\46} are reported in Table~\ref{molecular-table}.

\section{An overview of the radio emission from {\46}}
 \label{thor}
The so-called THOR+VGPS  image  of {\46}~SNR at 1.4~GHz is presented in Fig.~\ref{1.4thor}. 
The mapped region also shows at RA $\simeq$ \RA{19}{17}{28}, Dec $\simeq$ \Dec{11}{55}{45} (J2000) the \ion{H}{ii} region G046.495$-$00.241 reported in The \it WISE \rm Catalog of Galactic \ion{H}{ii} Regions \citep{and14}.\footnote{Alternative names used in the literature to refer to this thermal source are G46.5$-$0.2 \citep{kuchar1990} and G46.495$-$0.25 \citep{quireza06}.}

At radio wavelengths  {\46} consists of an almost circular shell of $\sim17^\prime$ in diameter, centred at RA $\simeq$ \RA{19}{18}{04}, Dec $\simeq$ \Dec{12}{09}{31}. The SNR surface brightness at 1~GHz is $\Sigma_{\mathrm{1~GHz}} \simeq 8.3 \times 10^{-21}$~W~m$^{-2}$~Hz$^{-1}$~sr$^{-1}$. Although $\sim$1700 times fainter than Cas~A (the brightest SNR in the Galaxy), {\46} still belongs to the large group of galactic SNRs with intermediate values of $\Sigma_{\mathrm{1~GHz}}$ \citep{green19}. 
The brightness distribution in {\46} is not uniform but it presents several spots and filamentary  structures throughout the SNR shell. The brightest parts of the remnant are localised towards two well differentiated regions in the north and the south of the shell comprising a sort of ``caps''. The former is approximately $10^\prime \times 3^\prime$ in size with two spots at its sides towards RA$\sim$\RA{19}{18}{07}, Dec$\sim$\Dec{12}{16}{42} and RA$\sim$\RA{19}{17}{45}, Dec$\sim$\Dec{12}{15}{05}. The latter is $\sim 5^\prime \times 2^\prime$ in size, with the brightest part towards RA$\sim$\RA{19}{18}{06}, Dec$\sim$\Dec{12}{02}{39}. The filaments are mainly detected in the southern half of the source and are aligned approximately in the south-west to north-east direction. A featureless band of weaker emission crosses the remnant from east to west in the northern half of the shell. 
Additionally, a slight flattening of the shell's boundary can be noted along the south-western border, as well as a small dent in the  north-west rim in coincidence with the western spot of the bright northern feature. These features could be suggesting a deceleration of the shock due to interaction with the surroundings. 
The correlations of the radio shell of G46.8$-$0.3 with the molecular component of the ISM will be addressed in Sect.~\ref{CO}.

\begin{figure}[ht!]
  \centering
  \includegraphics[width=0.47\textwidth]{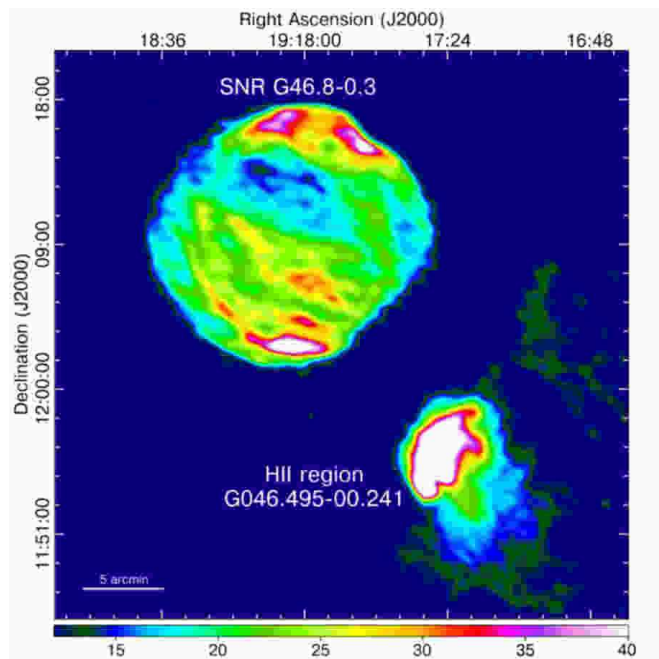}
  \caption{Radio view of {\46}~SNR at 1.4~GHz from combined THOR and VGPS continuum data \citep{beuther+16}. The colour scale varies linearly and is in units of mJy~beam$^{-1}$. The angular resolution is 25$^{\prime\prime}$ and the sensitivity is 0.72~mJy~beam$^{-1}$. The bright source at the southwest corner is the \ion{H}{ii} region named G046.495$-$00.241 in The \it WISE \rm Catalog of Galactic \ion{H}{ii} regions \citep{and14}.}
 \label{1.4thor}
\end{figure}

\subsection{Global radio continuum spectrum}
\label{global}
We have updated the integrated radio continuum spectrum of {\46}. For this purpose we measured flux densities over the whole extension of the source using publicly available images at low-radio frequencies from GLEAM, along with other at higher frequencies from THOR+VGPS, and GB6 (details of these observations are in Sect.~\ref{radio}), and the Radio Continuum Survey of the Galactic Plane 10~GHz made with the Nobeyama Radio Observatory (NRO, HPBW $\sim 2.^{\prime}7$, rms $\sim$ 33~mJy~beam$^{-1}$,  \citealt{handa87}). We also combined our new flux determinations with a carefully selected set of previously published estimates. While compiling these data, we rejected flux measurements previously reported with error estimates greater than 20\%, as well as data points showing a large scatter beyond the range of 2$\sigma$ best-fit values. 

In our study we also adopted the absolute flux density scale presented by \citet{per17} to tied fluxes for frequencies between 50~MHz and 50~GHz, the range  where the scale is accurate to 3\% and up to 5\% for measurements at the extreme frequency values. Table~\ref{fluxes} contains all the flux density values for {\46} covering about 2.5 decades in frequency from 30.9~MHz to 11.2~GHz. 
Only four fluxes in our list were not adjusted to this common scale because of the lack of information about the calibrator sources in the original publications. 

The integrated spectrum of {\46} at radio frequencies that we created is plotted in Fig.~\ref{spectrum}. The weighted best-fit to the data corresponds to a single power law slope $\alpha = -0.535 \pm 0.012$ ($S_{\nu} \propto \nu^{\alpha}$). This integrated spectral index is compatible with what is expected for a  middle-aged shell type  SNRs accelerating radio-emitting electrons via first-order Fermi mechanism. 
While comparing with previous determinations made for {\46}, our result is largely consistent with those by \citet{kovalenko-94-spec} ($\alpha =-0.53 \pm 0.10$), \citet{dubner96} ($\alpha \sim -0.53$), and \citet{sun11} ($\alpha=-0.54 \pm 0.02$), but it is flatter than the spectra measured by \citet{kas89s} ($\alpha \sim -0.6$) and \citet{taylor-92} ($\alpha = -0.72 \pm 0.14$, estimated using only flux densities values at 327~MHz and 4.8~GHz). 
We consider that our spectrum notably improves the previous ones published for {\46}, most of them with  larger relative error in the spectral index determination. This is essentially a consequence of the fact that we have filled gaps  along the radio frequency domain by compiling twice flux density values than \citet{kovalenko-94-spec} and \citet{sun11}, as well as that we have set the majority of the measurements to the most accurate absolute flux density scale presented to date. 

\begin{figure}[h!]
 \centering
\includegraphics[width=0.45\textwidth]{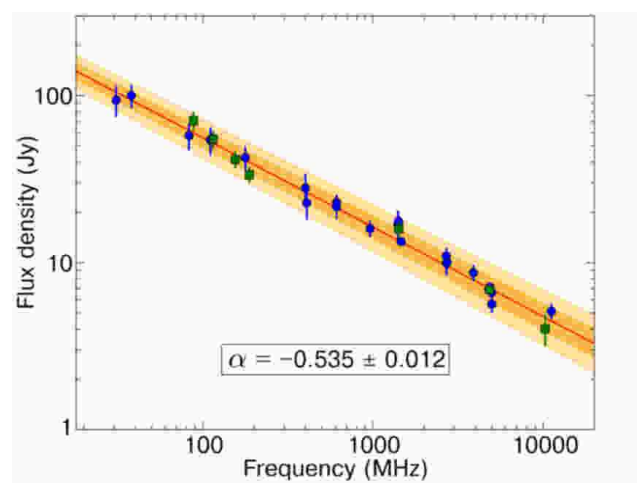}
\caption{Updated integrated radio continuum spectrum of SNR~\46 constructed using  fluxes listed in Table~\ref{fluxes}. Blue circles correspond to fluxes extracted from the literature, whereas the green squares are our new measurements obtained from radio surveys (see text for details). The solid line is the best-fitting curve to the weighted data and yields a spectral index $\alpha=-0.535 \pm 0.012$ ($S_{\nu} \propto \nu^{\alpha}$). Shaded regions represent the variation of the best-fit values at  1- and 2-$\sigma$ levels.}
 \label{spectrum}
\end{figure}

\begin{table}[h!]
\small\centering
\caption{Integrated flux densities on SNR \46 used to construct the radio continuum spectrum of the source presented in Fig.~\ref{spectrum}. Fluxes were set to the absolute scale of \citet{per17} (see text for details).}
\begin{tabular}{ccl}
\hline\hline
  Frequency   &  Scaled flux  &  \multirow{2}{*}{Reference} \\
  $[$MHz$]$   &     [Jy]      &                             \\\hline
  30.9 &	93.9 $\pm$ 18.8\tablefootmark{~$\ast$}          & \citet{kassim-88}         \\
    38 &  100 $\pm$	 15\tablefootmark{~$\ast$}          & \citet{holden-caswell-69} \\
    83 &	57.7 $\pm$	9.9	 & \citet{kovalenko-94}       \\
    88 &	70.9 $\pm$	8.9	 & This work (GLEAM)          \\
   111 &	54.1 $\pm$	9.8	 & \citet{kovalenko-94}       \\
   118 &	55.0 $\pm$	6.2	 & This work (GLEAM)          \\
   155 &	41.5 $\pm$	4.6	 & This work (GLEAM)          \\
   178 &	42.6 $\pm$	6.4	 & \citet{holden-caswell-69}  \\
   200 &	33.5 $\pm$	3.4	 & This work (GLEAM)          \\
   400 &	28.0 $\pm$	5.6\tablefootmark{~$\dagger$}       & \citet{downes-71} \\
   408 &	22.7 $\pm$	4.5	 & \citet{holden-caswell-69}  \\
   610 &	22.9 $\pm$	2.3\tablefootmark{~$\dagger$}       & \citet{moran-65}  \\
   610 &	21.5 $\pm$	3.2	 & \citet{holden-caswell-69}  \\
   960 &	16.0 $\pm$	1.6	 & \citet{trushkin-96}        \\
  1400 &	17.0 $\pm$	3.4\tablefootmark{~$\dagger$}       & \citet{downes-71} \\
  1400 &	16.7 $\pm$	2.5	 & \citet{holden-caswell-69}  \\
  1414 &	17.7 $\pm$	1.8	 & \citet{altenhoff-70}       \\
  1420 &	15.9 $\pm$	1.6	 & This work (THOR+VGPS)      \\
  1465 &	13.3 $\pm$	0.1	 & \citet{dubner96}           \\
  2695 &	11.0 $\pm$	1.1	 & \citet{altenhoff-70}       \\
  2700 &	 9.9 $\pm$	1.5	 & \citet{day-70}             \\
  2730 &	10.0 $\pm$	0.9	 & \citet{willis73}           \\
  3900 &	 8.7 $\pm$	0.9	 & \citet{trushkin-96}        \\
  4800 &	 6.9 $\pm$	0.2	 & \citet{sun11}              \\
  4850 &	 6.9 $\pm$	0.4    & \citet{condon-89}          \\
  4850 &	 7.2 $\pm$	0.4	 & This work (GB6)            \\ 
  5000 &	 6.6 $\pm$	0.7	 & \citet{caswell-clark-75}   \\
  5000 &	 5.6 $\pm$	0.6	 & \citet{angerhofer-77}      \\
 10300 &	 4.0 $\pm$	0.8	 & This work (NRO)            \\
 11200 &	 5.1 $\pm$	0.5	 & \citet{trushkin-96}        \\\hline
 \label{fluxes}
\end{tabular}
\tablefoot{
\tablefoottext{$\ast$}{No correction factor was applied to bring the measurement to the \citeapos{per17} scale because the corresponding frequency is outside its 50 MHz-50 GHz validity range.}\\ 
\tablefoottext{$\dagger$}{A correction to the absolute scale of \citet{per17} was not applied due to the lack of information on the flux density calibrator source.}}
\end{table}

\subsection{Local variations of the spectral index across \46}
\label{local}
We present the spectral index maps between 200~MHz and 1.4~GHz and between 1.4~GHz and 4.8~GHz in Figs.~\ref{fig-SPIX}a and 3b, respectively. The explanation of how these maps were constructed is presented in Sect.~\ref{radio}. Uncertainties for the local spectral index measurements on both maps are not higher than 0.1. We do not attempt to quantify the remaining parameter errors because we have no access to the collection of calibrated radio  visibilities to match the data at each frequency before constructing the spectral maps. Nevertheless, the spectral images we are presenting here are still useful in tracing local trends in the radio domain. Some contours of the $^{12}$CO ($J$ = 1-0) line emission are over-plotted on the spectral maps for reference.

\begin{figure*}
\centering 
\includegraphics[width=0.6\textwidth]{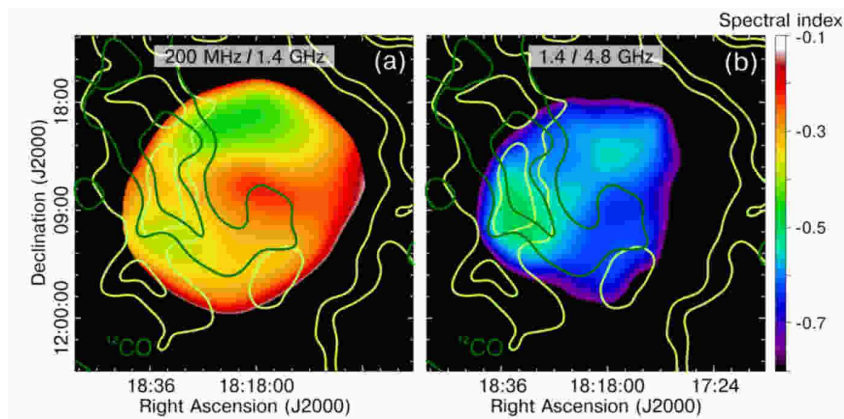}
\caption{Spectral index maps for {\46}~SNR constructed between \it (a) \rm 200 MHz and 1.4 GHz and \it (b) \rm 1.4 and 4.8~GHz. The images used are from GLEAM, THOR+VGPS, and GB6 surveys (see Sect.~\ref{radio}). Region with flux densities lower than 4$\sigma$ level of the respective rms noise in each image were clipped. The colour scale displayed to the right indicates the  spectral index measured (with a $4^{\prime}$ resolution) over the SNR at both frequencies bands. Dark and light green contours represent the intensity of the FUGIN $^{12}$CO ($J$ = 1-0) molecular line emission (at a  2$^{\prime}$ spatial resolution) integrated in the   30-50~km~s$^{-1}$ (levels: 500 and 800~K~{\kms}) and 50-66~km~s$^{-1}$ (levels: 1100 and 1400~K~{\kms}) ranges, respectively. A detailed treatment of the molecular features associated with {\46} is presented in Sect.~\ref{CO}.}
\label{fig-SPIX}
\end{figure*}

As revealed in Fig.~\ref{fig-SPIX}a, the distribution of spectral index values in the 200~MHz-1.4~GHz map is essentially flat, with a mean value $\alpha \sim -0.3$ over a horseshoe-like shape (in yellow-greenish shaded) that seems to align with the molecular material into which the SNR shock front is running, an issue we revisit in Sect.~\ref{CO}. The spectrum becomes slightly steeper ($\alpha \sim -0.45$) than towards the north portion of the remnant. By comparison, the spectral indices between 1.4 and 4.8~GHz look more uniformly distributed and steeper than at lower frequencies ranging from 200 to 1420~MHz. A moderate flatter spot ($\alpha \sim -0.5$) is also noticeable in the 1.4/4.8~GHz map towards the eastern side of \46. 

There seems to be in our spectral index maps a general trend of a slight concave-down behaviour (i.e., a steeper spectrum to higher radio frequencies). We recognise that the maps have not sufficient sensitivity to differentiate whether this signature  primarily occurs in the brightest portions of the remnant or not. If so, it might be a consequence of a higher effective shock compression undergone by low-frequency electrons in the interaction with adjacent clouds, which locally can increase the magnetic field strength and/or the particle density.
Even if this scenario is true for the case of {\46} it has not a measurable effect on the integrated spectrum of the source, which as shown in Sect.~\ref{global} it is well fitted by a single power-law (see Fig.~\ref{spectrum}).  
There are few cases of SNRs showing a steepening at high radio frequencies in their integrated spectra, with frequency cut-off varying from at around a few MHz up to GHz (e.g., Galactic SNRs S147, \citealt{xiao08} and HB~21, \citealt{pivato13}, and J0527-6549  in the Large Magellanic Cloud, \citealt{bozzetto2010}). All of them are evolved SNRs for which the observed global spectral form was explained in the standard diffusive acceleration shock theory by  the  compression of the Galactic magnetic field in dense interstellar regions combined or not with a possible contribution of synchrotron losses of the high-energy electrons (see \citealt{urosevic2014} and  \citealt{xiao08} for a review of this subject). 
However, we consider the occurrence  of synchrotron losses in \46 unlikely.
Indeed, the radiative cooling time of GeV electrons producing radio synchrotron emission in a magnetic field $B$ scales as 
$t_{\mathrm{syn}} \approx 1.3 \times 10^{10}\, (B/ \mathrm{\mu G})^{-2}\,(E/\mathrm{GeV})^{-1}$~yr \citep{gaisser1998} and in the case of  \46 it results to be $t_{\mathrm{syn}}\sim 5 \times 10^{6}$~yr. 
In this calculation we have assumed $E=3$~GeV for the energy of the electrons, along with an interstellar magnetic field strength in the clouds 
$B_{\mathrm{ISM}}\simeq n^{1/2}$~$\mu$G \citep{hollenbach1989} and $B=4B_{\mathrm{ISM}} \sim 30$~$\mu$G in the upstream region because of the shock compression in a gas with a mean molecular hydrogen of 50~cm$^{-3}$ (obtained from  Table~\ref{molecular-table} presented below in this paper). 
We notice that our $t_{\mathrm{syn}}$ estimate is absolutely comparable with typical values derived for SNRs evolving in relatively dense ambient medium (see, for instance, Fig.~5.3 in \citealt{aharonian2004}) but, as we shall show in Sect.~\ref{age}, it is much larger than the SNR's age ($t_{\mathrm{SNR}} \sim 1 \times 10^{4}$~yr). More sensitive radio observations towards {\46}, particularly at the low-frequency band, are required to confirm the validity of our analysis of local variations in the spectral index, before gaining a conclusive explanation.

\section{Results of the \ion{H}{i} data}
\label{HI}
Figure~\ref{HI-figure} displays the \ion{H}{i} intensity  in a velocity range from 10 to 66~km~s$^{-1}$ superposed on the 1.4~GHz radio continuum emission from {\46} and  the nearby \ion{H}{II} region G046.495$-$00.241.

The interstellar hydrogen appears to be distributed all around the rim of {\46} throughout the velocity interval presented in Fig.~\ref{HI-figure}. Over the SNR shell the brightness temperature of the \ion{H}{i} decreases significantly as compared with values observed outside the {\46} boundary. 
There are \ion{H}{i} local depressions, specially noticeable towards the northern and southern parts of the SNR shell. We have checked that the spectra extracted from these regions show clear signs of absorption. Due to their morphological correspondences with the brightness continuum emission from the remnant, we argue that these features correspond to absorbing atomic gas located in front of {\46} relative to the line of sight. There is also a faint  belt-like feature of \ion{H}{i}, seen for velocities greater than  $\sim$30~km~s$^{-1}$, crossing the remnant from northeast to northwest. We have not found signs of strong absorption (only a 3$\sigma$ signal) in the \ion{H}{i} profiles (not presented here) obtained in this region. 
Of course, the low brightness surface of the remnant could be limiting the absorption, but also it makes possible that most of the atomic gas forming the belt-like feature is placed on the far side of the SNR. In the southwest of the remnant the \ion{H}{i} is seen in absorption against the continuum radiation from the nearby  \ion{H}{ii} region G046.495$-$00.241. 

In summary, the distribution of the neutral gas towards {\46} does not show evidences for neighbouring structures that may have been impacted by the SN shock. In addition, no signatures for material swept-up due to the activity of the remnant or its progenitor are observed. If an expanding \ion{H}{I} structure were present it would be observed  in a reduced velocity interval of the atomic distribution as a region of low emission surrounded by a shell completely or partially correlated with the outer boundary of the SNR. An structure of this kind is not observed in the case of {\46}. Our analysis based on VGPS data confirms the previous result obtained by \citet{park+13} in direction to {\46} using lower spatial resolution (HPBW=3$^{\prime}$.35) 21~cm line data from the Inner-Galaxy Arecibo L-band Feed Array (I-GALFA) \ion{H}{i} survey \citep{gibson2012}.

\begin{figure*}[ht!]
\centering 
\includegraphics[width=0.8\textwidth]{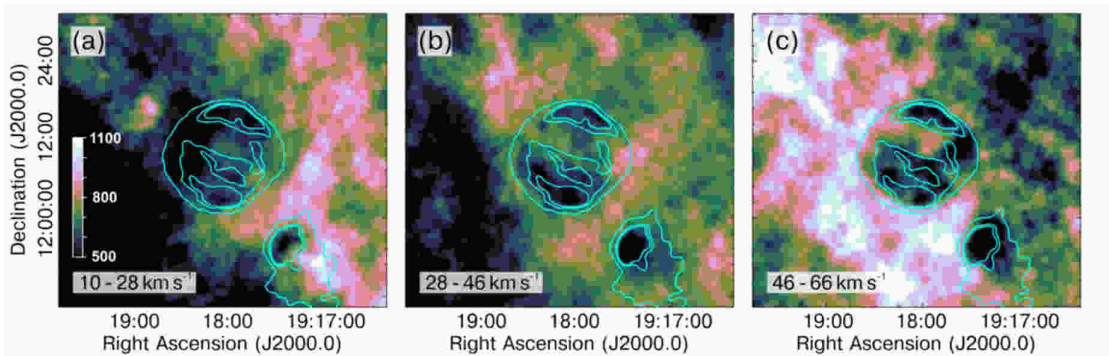} 
\caption{Velocity channel maps of the \ion{H}{i} emission in {\46}~SNR from  VGPS spectral line data, overplotted with the THOR+VGPS contours (levels: 70, 120, and 150~mJy~beam$^{-1}$) of the radio continuum radiation from the remnant and the \ion{H}{ii} region G046.495--00.241. 
For comparison, the radio continuum image was smoothed to the 60$^{\prime\prime}$ spatial resolution of the \ion{H}{i} data. Before constructing the maps presented in the figure, an appropriate mean background level was subtracted to each \ion{H}{i} velocity channel. The integration velocity range is indicated at the bottom left corner of each panel. All the velocity ranges are displayed with the linear colour scale in units of K~km~s$^{-1}$ shown in panel (a).}
\label{HI-figure}
\end{figure*}

\subsection{Newly derived distance for {\46}~SNR and the nearby \ion{H}{ii} region  G046.495--00.241}
\label{distances}
There is no agreement on the estimates for the distance to {\46} reported in the literature. \citet{sato1979} suggested a distance between 6.8 and 8.6~kpc from the analysis of \ion{H}{i} absorption. Later, \citet{ranasinghe2018-distance} from the analysis of \ion{H}{i} spectral features proposed lower and upper limits of the distance to be 5.7 and 11.4~kpc. More recently, \citet{liu2019} listed a distance of 6.4~kpc through the detection of a hydrogen radio recombination line (RRL) using data from The Survey of Ionized Gas of the Galaxy carried out with the Arecibo telescope (SIGGMA, \citealt{liu2013}).

In order to improve the distance estimate to our target and to get a complete insight into the physical association with the rich ambient medium which we discuss later in Sect.~\ref{CO}, we begin with a discussion of the properties of the \ion{H}{i} spectrum seen against the remnant.  To do so, we used the \ion{H}{i} cube from the VGPS. 
Figure~\ref{HIspectra}a shows \ion{H}{i} emission and absorption profiles for \46. The emission spectrum was obtained in the line of sight directly towards the bright southern portion of the source, while the absorption spectrum was constructed by subtracting the emission profile from a spectrum averaged over a set of regions close to the remnant. 
Firstly, we mention that the absence of any absorption feature at negative velocities in the spectrum\footnote{We recall there is no kinematic distance ambiguity for the \ion{H}{i} gas in the first quadrant of the Galaxy at negative velocities.} means that the SNR is inside the solar circle. Thus, according to the circular rotation curve model of the Galaxy by \citet{reid2014},  an upper limit on the distance of $\sim$11.4~kpc (which approximately corresponds to the radial velocity of approximately  0~km~s$^{-1}$) can be established. 
Secondly, we detect absorption up to the velocity of the tangent point, which at the SNR’s longitude is around 65~km~s$^{-1}$ \citep{reid2014}. This represents a lower limit on the distance of $\sim$5.7~kpc. The \ion{H}{i} absorption beyond the tangent point velocity probably corresponds to velocity perturbations near the tangent point, caused for instance by the SNR’s expansion. 
Additionally, we stress that significant and continuous absorption is observed from the velocity $\sim$~40.4~km~s$^{-1}$ to that of the tangent point, which might even imply that a better estimate of the minimum distance is the far side distance for $\sim$~40.4~km~s$^{-1}$. Therefore, we assign a kinematic distance $8.7 \pm 1.0$~kpc to the remnant. 
The error in our measurement is primarily caused by uncertainties in determining radial velocities along the line of sight, departures from circular motions of the Galaxy, as well as uncertainties associated with the selection of the rotation curve and solar motion. All of these contributions were added in quadrature in our calculations.

Now, we turn our attention to the \ion{H}{ii} region G046.495$-$00.241 seen in the area mapped in Fig.~\ref{1.4thor} as a bright source with an angular diameter of $\sim7^{\prime}$. 
The centre of this source and that of the remnant are separated from each other by approximately $\sim16^{\prime}.4$. 
In order to revisit the kinematic distance to G046.495$-$00.241, in Fig.~\ref{HIspectra}b we present absorption and emission profiles that we have constructed using spectral line VGPS observations directed to the central portion over the thermal source. 
The velocity for the \ion{H}{ii} region deduced in \citet{liu2019} from hydrogen RRL is 58.4~km~s$^{-1}$. The corresponding near and far distances inferred with the rotation curve of \citet{reid2014} are 4.2 and 7.3~kpc, respectively. 
As revealed in Fig.~\ref{HIspectra}b, the \ion{H}{i} spectrum exhibits absorption at the $\sim$4$\sigma$ level beyond the RRL velocity up to the velocity of the tangent point, which at the \ion{H}{ii} region's Galactic longitude is $\sim$65.6~km~s$^{-1}$.
Therefore, we placed the thermal source at the far distance of 7.3 $\pm$ 1.2~kpc, in accord with the distance that we established in Sect.~\ref{CO} to the molecular emission spatially coincident with it. 
We find that our distance determination  is markedly consistent with the 7.8~kpc value reported by  \citet{quireza06} from measurements of intensity and width in helium RRL data, but incompatible with those provided by \citet{kuchar1990} ($d=3.9$~kpc) obtained using \ion{H}{i} absorption spectrum with data from the Boston University-Arecibo Galactic \ion{H}{i} Survey, \citet{and09} ($d=3.8\pm 0.6$~kpc) based on a combination of \ion{H}{i} and CO sky surveys, and \citet{and14} ($d=4.7\pm 0.1$~kpc) in their \it WISE \rm census of \ion{H}{ii}.

\subsection{{\46}'s age estimate}
\label{age}
Taking into consideration the $8.7\pm1.0$~kpc distance to {\46} that we have determined in this work and its angular radius $\simeq$8$^{\prime}$.5 measured from the 1.4~GHz THOR+VGPS image, the physical radius of the SNR is approximately $R_\mathrm{SNR}\simeq21.5$~pc. We used this result to estimate the dynamical age of the remnant by employing evolutionary models. 
To do this, we notice that there is no appreciable optical emission in apparent association with {\46} in surveys such as VTSS\footnote{The Virginia Tech Spectral-Line Survey, \url{http://www1.phys.vt.edu/~halpha/\#Images}.} \citep{dennison+98}, DSS,\footnote{The STScI Digitized Sky Survey, \url{https://archive.stsci.edu/cgi-bin/dss_form}.} APASS,\footnote{The AAVSO Photometric All-Sky Survey, \url{https://www.aavso.org/apass}.} or IPHAS\footnote{The INT Photometric H$\alpha$ Survey of the Northern Galactic Plane, \url{https://www.iphas.org/}.} \citep{barentsen+14}, indicative of radiative shocks. 
Also, no X-ray emission was detected associated with the SNR in the only observation ($\sim$9.9~ks) for this remnant available in the \textit{Chandra} Data Archive.\footnote{The non-detection of X-ray emission would support that the SNR has significantly cooled.} 
Nevertheless, although it is not a strong constrain to the age, the large size of {\46} could suggest that it is an evolved object. Under this interpretation, we hypothesise that the remnant is in the later adiabatic expansion stage of its evolution. We caution that, although it can be reliable in terms of the order of magnitude, our modest estimation of the {\46}’s age suffers from large uncertainties derived from the assumed parameters values and the assumption of the stage in the SNR’s evolution.

According to the numerical approach of \citet{cox1972}, for a Sedov-Taylor (ST)  blast wave shock with a specific heat ratio $\gamma=5/3$, the age for {\46} given by $t_\mathrm{SNR}= 14.5 \, (R_\mathrm{SNR})^{5/2} \, ({n_0}/E_{51})^{1/2}$ is about $t_\mathrm{SNR}^\mathrm{ST}\sim 1 \times 10^4$~yr. In the calculations above, $E_{51}$ is the kinetic energy of the explosion normalised to the canonical value $10^{51}$~erg, assumed to be $E_{51}$ = 1 in this work, \footnote{By applying spherically symmetric SNR evolution models to a sample of 43 objects in the Galaxy with X-ray observations and reliable distance estimates, \citet{leahy2020} calculated a mean value $\sim$$4\times10^{50}$~erg for the energy released in a SN event.} 
while $n_0$ is the pre-shock number density for which we adopted $n_0\sim 0.1$~cm$^{-3}$, a value of reference between a typical density inside a wind bubble excavated by the stellar progenitor ($n_{0}\sim 0.01$~cm$^{-3}$) and that of the diffuse intercloud gas ($n_0\sim 1$~cm$^{-3}$) \citep{inoue2012}. 
In addition, the expansion velocity of the SN shock during the adiabatic phase is $v_{\mathrm{s}}\approx 13.4 \times 10^{4}\, (t_{\mathrm{SNR}}^{ST})^{-3/5}\,(E_{51}/n_{0})^{1/5}$~km~s$^{-1}$ $\sim 845$~km~s$^{-1}$. Following the approach presented by \citet{blondin1998}, the transition time to the radiative phase will occur at $t_{\mathrm{tr}}\approx 2.9\times 10^{4}\,E^{4/17}_{51}\,n^{-9/17}_{0}$~yr $\sim 1 \times 10^{5}$~yr, with a velocity of the shock wave  $v_{\mathrm{s}}\approx 260\, E_{51}^{1/17} \, n_{0}^{2/17}$~km~s$^{-1}$$\sim 200$~km~s$^{-1}$.

\begin{figure}
  \centering
 \includegraphics[width=0.45\textwidth]{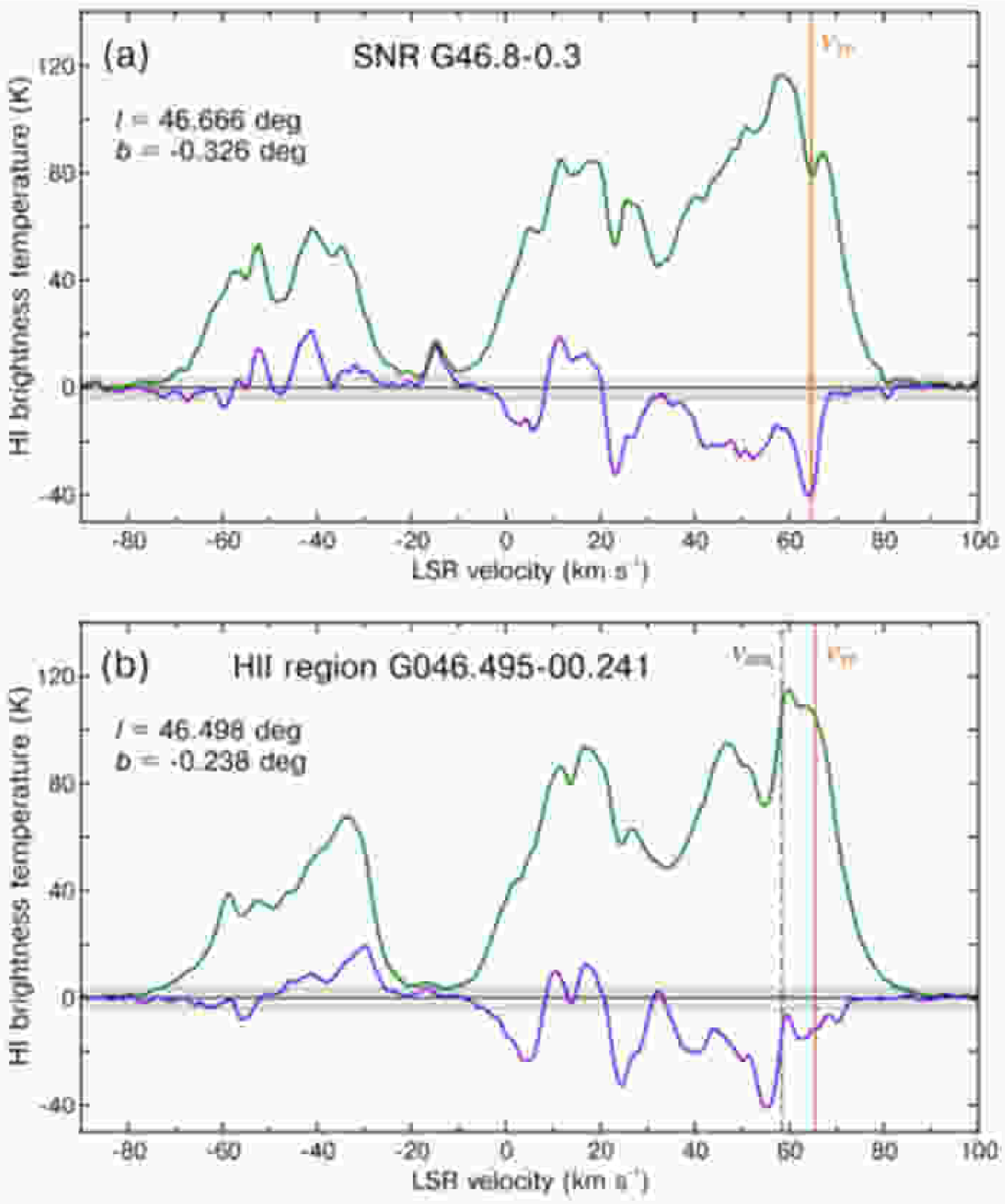}
 \caption{\ion{H}{i} line emission (green curve) and absorption (violet curve) spectra against \it (a) \rm G46.8$-$0.3~SNR and \it (b) \rm  G046.495$-$00.241 \ion{H}{ii} region. The solid  vertical line in each panel marks the velocity of the tangent point at the longitude of the sources according to the \citet{reid2014} Galactic rotation curve, whereas the dashed vertical line in panel \it (b) \rm is for the hydrogen RRL at 58.4~km~s$^{-1}$ from G046.495$-$00.241 recorded in \citet{liu2019}. The horizontal shaded regions indicate the rms noise level of $\sim$4~K in the \ion{H}{i} absorption spectra.}
  \label{HIspectra}
\end{figure}

\section{Results of the CO data}
\label{CO}
\subsection{Morphology of the CO molecular gas components}
\label{COmorphology}
Figure~\ref{COaverage} depicts the average spectrum of the $^{12}$CO and $^{13}$CO ($J$ = 1-0) molecular lines over a circular region of radius 14{\amin} concentric with the remnant \46. Multiple velocity components are evident towards the SNR between $\sim$10-65~km~s$^{-1}$. In what follows, we investigate the spatial distribution of the molecular gas around \46 traced by CO isotopologues, leaving the discussion on the spectral characteristics of the molecular ambient matter to Sect.~\ref{kinematic}. We have omitted the analysis of the molecular gas responsible for the very narrow (FWHM $\mathrm{\Delta v \simeq 2.5}$~km~s$^{-1}$) emission feature at $\sim7$~km~s$^{-1}$, because we have not found any evidence of association with the SNR.

\begin{figure}[h!]
\centering
\includegraphics[width=0.45\textwidth]{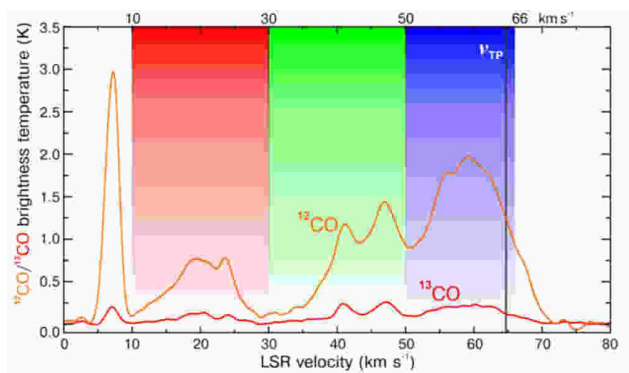}
\caption{Overall CO spectra covering the SNR~\46 region. The three shaded areas correspond to LSR velocity ranges for which there is good spatial correspondences between the molecular matter and the remnant (see discussion in the text). The velocity of the tangent point ($\sim$65~km~s$^{-1}$, \citealt{reid2014}) is marked with a black vertical line.}
\label{COaverage}
\end{figure}

In Fig.~\ref{CO-figure} we present three-colour integrated intensity maps of the molecular environments in direction to \46 ($^{12}$CO $J$ = 1-0 in red and $^{13}$CO $J$ = 1-0 in green) obtained by FUGIN, together with the SNR radio synchrotron emission at 1.4~GHz (in blue) as traced by THOR+VGPS data. In such a representation  $^{12}$CO observations typically delineate emission mostly coming from low density gas in the outer layers of the cloud, while the $^{13}$CO data generally trace optically thin emission from denser regions. The mapped CO emissions correspond to molecular structures for which we consider there is a convincing match with the remnant. They are in three velocity intervals: 
10-30~km~s$^{-1}$, 
30-50~km~s$^{-1}$, and 
50-66~km~s$^{-1}$. 
The molecular matter in these velocity intervals is hereafter referred to as Regions A, B, and C, respectively. According to their kinematical properties, which are analysed in detail in Sect.~\ref{kinematic}, the regions are also separated in subareas or clouds. In the notation we have adopted throughout this work,\footnote{Our notation is similar to that adopted by \citet{sano-17-RCW86} in their study of RCW~86~SNR.} each cloud is labelled with a capital letter (A, B, or C), a number referencing the distance to the cloud, and the N, E, W, SW,  or C letter depending on whether the cloud is located towards the north, east, west, southwest or centre of the remnant, while the \ion{H}{ii} abbreviation is added for the gas immersed in the G046.495$-$00.241 region.

\begin{figure}[h!]
\centering 
\includegraphics[width=0.45\textwidth]{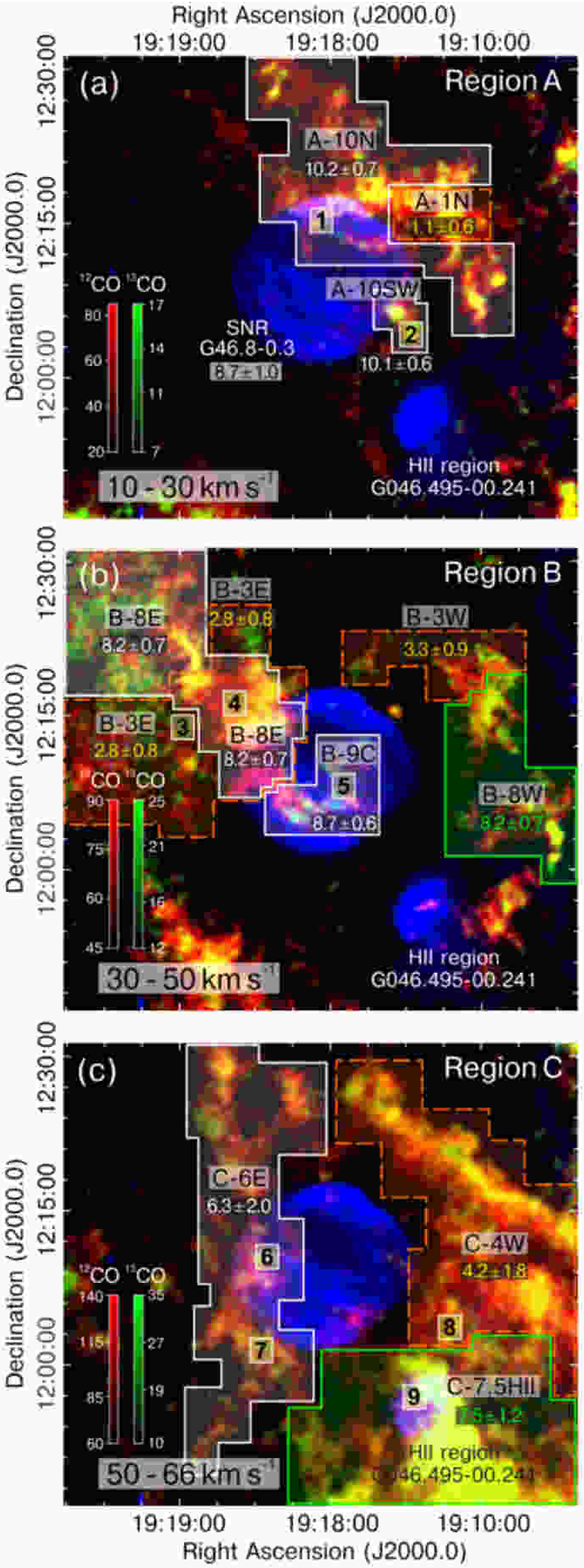}
\caption{Colour representation of the molecular ambient towards {\46}~SNR, traced by the \element[][12]{CO} $J$ = 1-0 (in red) and the \element[][13]{CO} $J$ = 1-0 (in green) rotational line emissions. Yellow regions are areas where both CO isotope emissions overlap. The SNR radio continuum radiation at 1.4~GHz from THOR+VGPS surveys is represented in blue. The velocity range used to integrate the CO intensity is shown at the bottom left corner of each panel. The name of the discovered molecular features and their kinematic distances are also labelled. White continuous polygons enclose structures associated with {\46}. Non associated clouds localised at the far distance of their central velocity are encircled by green continuous polygons, whereas foreground molecular gas along the line of sight is inside of orange dashed polygons. Numbers from 1 to 9 spread out in panels \it a\rm-\it c \rm indicate the test-regions used to construct the spectra presented in Fig.~\ref{COspectra}.}
\label{CO-figure}
\end{figure}

As shown in Fig.~\ref{CO-figure}a, the gas in Region~A at velocities in the  10-30~km~s$^{-1}$ range comprises two molecular components distributed around the SNR shell. They are: \it (i) \rm the cloud named A-10N  and \it (ii) \rm the cloud A-10SW. The former follows the northern edge of the remnant  impressively. Besides, part of this gas appears  to be projectionally embedded in the synchrotron emission from the remnant encompassing a distortion observed in the outer envelope of the SNR’s shell at the position RA $\sim$ \RA{19}{17}{45}, Dec $\sim$ \Dec{12}{15}{40}. This correspondence is compatible with a shock wave of the SNR interacting with the ambient matter. The molecular concentration also extends outside the northern limb of the remnant, towards the southwest direction. The second molecular component, labelled A-10SW, consists of a set of clumps concentrated on the southwest edge of the SNR, exactly at the site where the shock front is flattened. 

For intermediate velocities, 30-50~km~s$^{-1}$, the ambient medium in Region~B can be divided into three main zones (Fig.~\ref{CO-figure}b): \it (i) \rm  the molecular concentration towards the east, including the clouds  B-3E and B-8E, which correlates the remnant's limb and extends out about $20^{\prime}$ in the northeast direction beyond the shock front; \it (ii) \rm the molecular gas distributed in the interior of the SNR’s shell along the northeast-southwest orientation, labelled cloud B-9C, which appears to pass through bright patches of the radio continuum emission; and \it (iii) \rm the western region comprising clouds B-3W and B-8W, which seems to be part of a molecular wall in the environment of our target source.

At the highest velocities, 50-66~km~s$^{-1}$, there is molecular gas seen in projection all around the remnant (Fig.~\ref{CO-figure}c). The best match is observed towards the east where the CO in the C-6E region closely follows the edge of {\46}. 
The spatial correlation with the remnant decreases to the western half of the map, where the cloud C-4W is observed outside the SNR shock front, but still around it. 
Towards the southwest is the  molecular gas concentration named C-7.5HII, probably the birthplace of the star-forming \ion{H}{ii} region G046.495$-$00.241. 

Figure~\ref{CO-figure} also reveals that all the identified molecular regions contain multiple enhancements where the $^{13}$CO $J$ = 1-0 line emission is detected at levels higher than 5$\sigma$ in coincidence with local maxima in the $^{12}$CO emission. 
The $^{13}$CO excess noticeable in the clouds A-10N, A-10SW, B-8E, and B-9C seen in projection onto the {\46}'s radio shell may indicate strong compression of the molecular gas due to interaction with  the SNR shock front.

At a first glimpse {\46} seems to be embedded within a dense molecular cavity. The existence of an almost complete CO-line cavity at 52~km~s$^{-1}$ associated with the remnant has been already proposed by \citet{sofue2021} based on FUGIN survey of $^{12}$CO- and $^{13}$CO ($J$ = 1-0)-line channel maps, the same data  we used in the current work. These authors did not determined the kinematic distance to such a structure but mentioned possible near and far values of 3.9 and 7.1~kpc, respectively. 
On the basis of an CO-\ion{H}{I} combined spectral analysis, in next section we determine the distance to each molecular structure unveiled in our work and hence we revisit the hypothesis of a cavity in the molecular medium around \46.

\subsection{Kinematics of the CO molecular gas components}
\label{kinematic}
Here, we focus on determining the kinematic distances to the uncovered  molecular structures comprising  regions A-C shown in Fig.~\ref{CO-figure}. To achieve this we have extracted  multiple \ion{H}{i} and CO emission/absorption spectra covering the complete extension of each cloud, some of them are presented in Fig.~\ref{COspectra}. The test-areas corresponding to our examples are marked with numbers from 1 to 9 in Fig.~\ref{CO-figure}. In Table~\ref{molecular-table} we report averaged physical properties (peak intensity, velocity, line-width, angular size, column density, mass, and number density) computed over each CO structure unveiled in this work. Column densities and masses were calculated from the expressions presented in Sect.~\ref{ism} and using the distance estimates provided here.

The sources of uncertainty in our distance results computed from CO-\ion{H}{i} combined spectra are the same as those of our distance determination to the SNR and the \ion{H}{ii}  region derived in Sect.~\ref{distances} from \ion{H}{i}  profiles only. As listed in Table~\ref{molecular-table}, the final error in the measured distances to the molecular clouds are better than 10\%, with the exception of those located at the near-side or close to the tangent point, for which the mean error is $\sim$33\%. On the other side, the final error in the measured distances to the molecular clouds combined with both the error of at least 30\% in the CO-to-H2 conversion factor \citep{bolatto2013} and  the uncertainties in isolating overlapping regions of sky, translate into errors in the calculated properties (i.e. column densities of each velocity component and mass)  of the interstellar gas. See further details in  Table~\ref{molecular-table}.  

\begin{table*}[h!]
\small
\centering
\caption{
Measured and derived properties for the molecular structures identified in direction to {\46}~SNR. The reported values are relative to the \element[][12]{CO} $J$=1-0 line transition and represent averages obtained over the entire extension of each individual cloud.
The detailed analysis of the spectral characteristics of the molecular gas is presented in Sect.~\ref{kinematic}.}
\label{molecular-table}
\begin{tabular}{c l c c c c c c c c c}\hline\hline
(1) & ~~~~~~(2) & (3) & (4) & (5) & (6) & (7) & (8) & (9) & (10) & (11)\\
 Velocity                                     &%
 Molecular                                    &%
 \multirow{2}{*}{Distance}                    &%
 \multirow{2}{*}{$T_\mathrm{peak}$}           &%
 \multirow{2}{*}{$v_\mathrm{LSR}$}            &%
 \multirow{2}{*}{$\Delta v_{\mathrm{FWHM}}$}  &%
 \multirow{2}{*}{Size}                        &%
 \multirow{2}{*}{$N_{\mathrm{H}_2}$}          &%
 \multirow{2}{*}{$M_\mathrm{tot}$}            &%
 \multirow{2}{*}{$n_{\mathrm{H_2}}$}          &%
SNR/MC \\
 range & ~~~~cloud  & & & & & & & & & association \\
 $[\mathrm{km~s}^{-1}]$ & ~~~~(MC) &%
 [kpc] &%
 [K] &%
 [km~s$^{-1}$] &%
 [km~s$^{-1}$] &%
 [arcmin$^{2}$] &%
 [$10^{21}$~\cm{-2}] &%
 [$10^3~\mathrm{M}_\sun$] &%
 [cm$^{-3}$]\\\hline
 \multirow{3}{*}{10 - 30}  & A-1\,N   &  1.1 $\pm$ 0.6  & 5.7   & 15.5   &  6.9   &  8 $\times$ 5   & 4.7  & 0.4   &  760 &  no  \\
                           & A-10\,N  & 10.2 $\pm$ 0.7  & 3.6   & 18.7   &  6.5   & 22 $\times$ 10  & 3.3  &  214  &   23 &  yes \\
                           & A-10SW   & 10.1 $\pm$ 0.6  & 4.3   & 19.3   &  6.7   &  7 $\times$ 4   & 2.7  &   13  &   55 &  yes \\\hline
 \multirow{5}{*}{30 - 50}  & B-3\,E   &  2.8 $\pm$ 0.8  & 3.5   & 41.0   &  5.3   & 19 $\times$ 11  & 2.7  &   10  &   70 &  no  \\
                           & B-8\,E   &  8.2 $\pm$ 0.7  & 3.3   & 47.1   &  4.3   & 21 $\times$ 7   & 5.4  &  106  &   52 &  yes \\
                           & B-9\,C   &  8.7 $\pm$ 0.6  & 3.2   & 40.8   &  4.5   &  9 $\times$ 7   & 2.5  &   23  &   40 &  yes \\
                           & B-3\,W   &  3.3 $\pm$ 0.9  & 4.8   & 48.2   &  3.8   &  7 $\times$ 5   & 3.1  &  2.7  &  175 &  no  \\
                           & B-8\,W   &  8.2 $\pm$ 0.7  & 5.3   & 46.9   &  6.0   & 15 $\times$ 10  & 4.7  &  107  &   50 &  no  \\\hline
 \multirow{3}{*}{50 - 66}  & C-6\,E     &  6.3 $\pm$ 2.0  & 4.5  & 63.4  &  7.5   & 33 $\times$ 8   & 7.4  &  160  &   65 &  uncertain \\
                           & C-4\,W     &  4.2 $\pm$ 1.8  & 4.9  & 57.5  & 10.6   & 20 $\times$ 9   & 9.6  &   90  &  180 &  no  \\
                           & C-7.5\,HII &  7.5 $\pm$ 1.2  & 5.3  & 56.1  &  8.8   & 27 $\times$ 7   & 9.8  &  255  &   86 &  no  \\\hline\hline
\end{tabular}
\tablefoot{
Cols.~(1)-(2): Velocity intervals and names corresponding to the CO clouds identified in this work. For the sake of reference see Fig.~\ref{CO-figure}. 
Col.~(3): Kinematic distances to each cloud computed from CO and \ion{H}{i} line emission/absorption spectra. The atomic and molecular data come from the VGPS and FUGIN surveys, respectively.
Cols.~(4)-(6): Peak intensity ($T_{\mathrm{peak}}$), line-of-sight velocity ($v_{\mathrm{LSR}}$), and line-width (full-width-half-maximum, $\Delta v_{\mathrm{FWHM}}$), respectively. 
Velocity measurements have a mean associated error of 7.6~km~s$^{-1}$, as explained in Sect.~\ref{ism}. The reported estimates were calculated by fitting Gaussian functions to the CO-\ion{H}{I} spectra. 
Col.~(7): Average dimensions (height $\times$ width) of each CO cloud. 
Cols.~(8)-(9): Molecular hydrogen column density ($N_{\mathrm{H_{2}}}$) and total mass ($M_{\mathrm{tot}}$) of the clouds.  Both magnitudes were calculated using expressions given in Sect.~\ref{ism}. The mean percentage error in our estimates of $N_{\mathrm{H_2}}$ and $M_\mathrm{tot}$ are, respectively,  $\sim$35\% and $\sim$50\% (for the interacting clouds); see discussion in Sect.~\ref{kinematic}. 
Col.~(10): Number density of the molecular hydrogen computed from the expression $n_{\mathrm{H_2}} = N_{\mathrm{H_2}}/L$ by assuming a path length along the line of sight, $L$, equivalent to the mean linear size of each cloud listed in Col.~(7). The error in $n_{\mathrm{H_2}}$ is typically 40\%. 
Col.~(11): Classification of the physical association between the {\46}~SNR and the discovered CO clouds, based on the analysis presented throughout the Sect.~\ref{CO} of this work.}
\end{table*}

\begin{figure*}
 \centering
\includegraphics[width=0.95\textwidth]{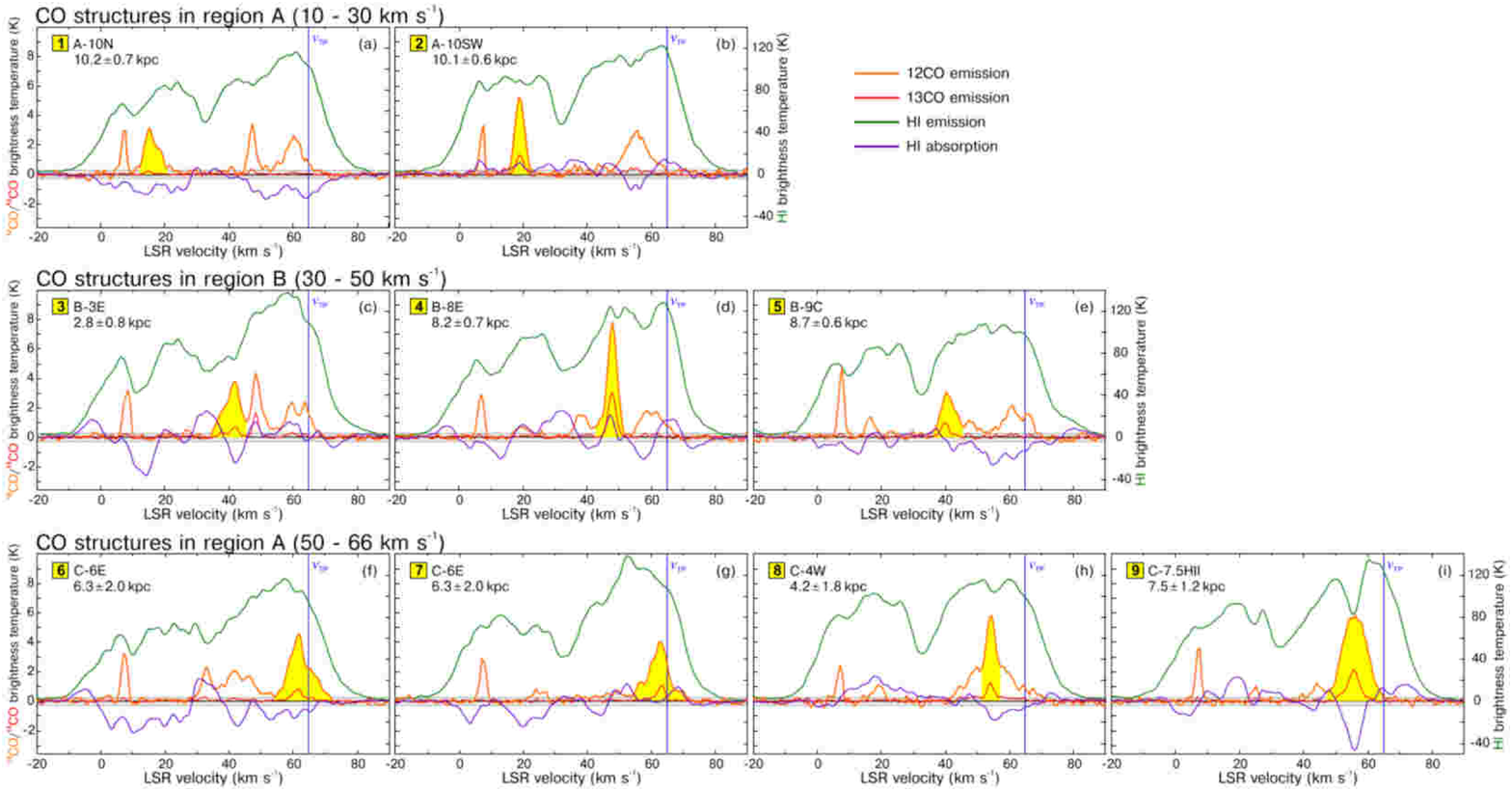}
\caption{A collection of composite spectra of the CO ($J$=1-0) and \ion{H}{i} 21~cm line emissions towards test-areas over the molecular structures discovered in the direction of SNR~\46, as shown in Fig.~\ref{CO-figure}. The correspondences between the molecular and atomic emissions were used to determine the kinematic distance to individual clouds (see analysis in Sect.~\ref{kinematic}). 
The shaded zones in yellow highlight the CO peak and the LSR velocity in the region we are interested in. The blue vertical line in each spectrum indicates the velocity of the tangent point according to the rotation curve model of \citet{reid2014}. For the remaining spectra (not shown here) used in our analysis, we have found in each region approximately the same behaviour exhibited in those presented here.}
\label{COspectra}
\end{figure*}

\subsubsection{Region~A: 10-30~km~s$^{-1}$}
Now, we analyse the molecular gas inside Region~A, spatially correlated with the northern and the southwestern edges of \46 (Fig.~\ref{CO-figure}a). 
The spectra corresponding to the  selected areas over this region appear in Figs.~\ref{COspectra}a-b. The excellent correspondence between the southern part  of the molecular gas emission from the cloud named A-10N and the 1.4~GHz brightest emission from {\46} is consistent with an interaction zone between the remnant and its immediate environment. 
If so, the distance to the molecular region should agree with that established for the SNR ($8.7\pm1.0$~kpc, see Sect.~\ref{distances}). The $^{12}$CO spectrum over the A-10N cloud seen (in projection) immersed in the radio shell of the SNR exhibits  a peak at $\sim15$~km~s$^{-1}$ with a $^{13}$CO counterpart. 
Using the flat rotation curve model of \citet{reid2014}, the kinematic distances associated with this velocity are 1.0 and 10.2~kpc. In addition, the \ion{H}{i} 21~cm profile in the molecular component exhibits dips corresponding to CO emission lines at velocities up to the velocity of the tangent point, the maximum velocity that can have the foreground clouds. 
The observed CO-\ion{H}{i} anti-correlation can be explained in terms of foreground molecular clouds along the line of sight absorbing the continuum radiation from the SNR \citep{roman+09}. Therefore, we assigned to the investigated cloud the far side distance of $10.2 \pm 0.7$~kpc. 
If the cloud were located at the near position of its velocity, absorbed features will be observed in the \ion{H}{i} profile correlated with CO emission lines up to the velocity of the clump only. 
We also notice the  $^{12}$CO line profiles in Region~A deviates to higher velocities relative to the $^{13}$CO line maximum  intensity. It is not surprising, as the latter is a  more faithful tracer of the denser regions in the molecular gas. At this stage, we remark that the spectra (not presented here) constructed over the molecular gas extending beyond the SNR boundary confirm that the A-10N cloud is a single structure placed at $\sim$10~kpc. 
Therefore, taking into consideration that we have established the SNR's location at $8.7 \pm 1.0$~kpc, we interpret the asymmetry in the $^{12}$CO line emission (line FWHM $\Delta v \simeq6.5$~km~s$^{-1}$) as a sign of perturbation caused by the SNR shock wave penetrating into the most massive ($\sim$$214\times10^3$~$M_{\odot}$) cloud interacting with {\46}, physically located at the backside of the remnant. The striking morphological correspondence of the molecular gas with the radio boundary also further supports the shock-cloud interaction idea.

In our inspection of the molecular gas in Region~A, we noticed that the spectra constructed over the western extreme of the northern cloud, where it curves to the south (at RA $\sim$ \RA{19}{17}{15}, Dec $\sim$ \Dec{12}{15}{20}, see Fig.~\ref{CO-figure}a), display a CO emission peak at $\sim 16$~km~s$^{-1}$ correlated with \ion{H}{i} self-absorption against warm 21~cm backgrounds (emitting at the same velocity as of the cloud). This behaviour implies that this part of the surveyed Region~A, referred to as A-1N,  is at the near distance of $1.1\pm0.6$~kpc. 
We estimated that the mass of this gas only represents a small fraction (< 1\%) of the molecular material in the north brightening in the velocity range 10-30~km~s$^{-1}$.

As shown in Fig.~\ref{CO-figure}a, the molecular gas in Region~A also comprises bright clumps in the southwest (cloud A-10SW), where the rim of \46 clearly deviates from a circular geometry. The CO profile presented in Fig.~\ref{COspectra}b was obtained in a test-area just outside the SNR over the clump with a central velocity of $\sim19$~km~s$^{-1}$. The correlation of the CO peak with a maximum in the \ion{H}{i} emission indicates that this part of the molecular region resides at the far position of $10.1 \pm 0.6$~kpc.

\subsubsection{Region~B: 30-50~km~s$^{-1}$}
Over the CO material brightening between 30 and 50~km~s$^{-1}$ in the eastern area outside the remnant we have found spectra showing \ion{H}{i} self-absorption features at $\sim$41~km~s$^{-1}$  (Fig.~\ref{COspectra}c) and other ones revealing CO peak intensities at $\sim$48~km~s$^{-1}$ in coincidence with \ion{H}{i} maxima (Fig.~\ref{COspectra}d). 
This means that the ambient matter eastwards of {\46} results of a chance coincidence of near-side (cloud B-3E at 2.8 $\pm$ 0.8~kpc), and far-side (cloud B-8E at 8.2 $\pm$ 0.7~kpc) components superposed along the line of sight. At RA $\sim$ \RA{19}{18}{30}, Dec $\sim$ \Dec{12}{13}{00} the molecular emission bends to the west and overlaps the eastern portion of {\46}. The spectral properties over this zone (spectra not included in Fig.~\ref{COspectra}) indicate it is an extension of the molecular material outside {\46}. Indeed, for the near-side gas the cold \ion{H}{i} embedded in the foreground molecular clouds absorbs the continuum radiation from the SNR and produces spectral \ion{H}{i} 21~cm absorption lines up to a velocity around 41~km~s$^{-1}$ of the material we are interested in, while over the far-side component \ion{H}{i} 21~cm valleys corresponding to CO emission lines features are observed up to the velocity of the tangent point. 
We have also estimated that the amount of gas in the near-side component constitutes a $\sim$10\% of the total material in the eastern region shown in Fig.~\ref{CO-figure}b. 
Conversely to the accidentally near- and far-side overlapping of the CO structures in the east of the SNR, the spectral properties on the central B-9C cloud that appears embedded in the SNR shell (RA $\sim$ \RA{19}{18}{00}, Dec $\sim$ \Dec{12}{06}{10}) at a velocity of about 40~km~s$^{-1}$ allow us to conclude that it corresponds to a single cloud   at the far kinematic distance of $8.7 \pm 0.6$~kpc. As an example, Fig.~\ref{COspectra}e presents a spectrum extracted over this central area. The obtained distance to the cloud is fully compatible with that we measured for {\46} and hence  we interpret  the enhanced filaments of radio continuum emission as a trace of the SNR shock impacting on the relatively dense ($\sim$40~cm$^{-3}$) B-9C cloud. It is worth mentioning that we have found little evidence of disturbance in the profiles of the CO extracted over the molecular gas superimposed to the remnant. This result, however, does not rule out association with {\46}.  

Finally, we mention that near and far molecular components with line-of-sight coincidences were detected westwards {\46} at a velocity of $\sim$48~km~s$^{-1}$ (spectra not shown here). These cloud components are inside the regions we have denoted B-3W and B-8W, respectively. As shown in Fig.~\ref{CO-figure}b, this ambient medium was not overtaken by the SNR shock. We have  determined that approximately only 3\% of the surveyed gas in the west is at  the near position of $3.3\pm0.9$~kpc forming the cloud B-3W, while the remaining material places at $8.2\pm0.7$~kpc inside the cloud B-8W (see Table~\ref{molecular-table}).

\subsubsection{Region~C: 50-66~km~s$^{-1}$}
\label{regionC}
To analyse the molecular material in Region~C we have divided it into three areas differing in their spectral properties. The first zone comprises the gas forming the large eastwards structure (cloud C-6E). Towards the western half of the field  is the second molecular component (namely C-4W zone). In this region, distanced up to $\sim$18{\amin} from {\46}, the CO ($J$=1-0) line emissions are significantly detected from RA $\sim$ \RA{19}{17}{45}, Dec $\sim$ \Dec{12}{27}{00} to RA $\sim$ \RA{19}{10}{00}, Dec $\sim$ \Dec{12}{00}{00} (see Fig.~\ref{CO-figure}c). 
Lastly, the third important area corresponds to the gas seen projected onto the G046.495$-$00.241 \ion{H}{ii} region (the C-7.5HII cloud). Before describing the spectra, we point out that because of the small velocity difference measured in the profiles of the C-6E and C-4W areas between their CO peaks and the tangent point ($v_{\mathrm{TP}}\sim$65~km~s$^{-1}$, derived from the \citealt{reid2014} model) it is not simple to determine conclusively whether the clouds are at the near or the far position. Even though  arguments based on morphological correlations and spectral correspondences between the CO and \ion{H}{i} gases indeed may help to estimate their locations along the line of sight, the uncertainties are significantly larger than those associated with the A and B clouds. Therefore, our interpretation of the spectral features obtained over the east and west areas in Region~C should be regarded with care. As we shall discuss below, the situation is different for the C-7.5HII zone because of the \ion{H}{ii} region --whose distance was determined in Sect.~\ref{distances}-- sits inside the molecular gas.

The CO profiles over the portion of the cloud C-6E superposed on the SNR shell show maxima at around 64~km~s$^{-1}$ correlated with \ion{H}{i} absorption dips. This spectral condition (observed practically at the velocity of the tangent point) could indicate that the molecular emission occurs at the kinematic distance $6.3\pm2.0$~kpc associated with the mean central velocity measured in the region. This estimate is, within uncertainties, in reasonable agreement with the distance to {\46}. 
A key point to note is that many of the spectra we constructed over the narrow portion of the C-6E cloud superposed on {\46} were found to be broadened (FWHM line $\Delta v$ $\sim$ 9~km~s$^{-1}$, see example in Fig.~\ref{COspectra}f). Analogous spectral features, which have been interpreted as the result of turbulent motions in the shocked gas, were found in other known SNRs-molecular cloud interactions (e.g., W44, \citealt{seta+04}; 
G347.3--0.5, \citealt{moriguchi+05}; 
Kes~69, \citealt{zhou+09}; 
G357.7+0.3, \citealt{rho+17}; 
HB~3, \citealt{rho+21}). 
We also found a similar line broadening in the CO spectra constructed over the  portion of the C-6E cloud immediately outside the SNR forward shock. 
In this case, the spectra exhibit strong CO emission lines with peak intensities at around $\sim$63~km~s$^{-1}$ without associated \ion{H}{i} self-absorption features (see the representative profiles in Fig.~\ref{COspectra}g). Therefore, from our spectral analysis on the C-6E region we concluded that it is a single cloud located at $6.3\pm2.0$~kpc. Taking into account the distance we measured to {\46} and the large uncertainties in our distance determination to the molecular gas in the C-6E zone, we cannot completely discard the possibility that the cloud is located in the immediate foreground relative to the SNR.  The good spatial agreement of C-6E with the outer part of the remnant may be considered an additional evidence for a possible physical relationship between them.

Notoriously, we found that the spectral properties over the western part of Region~C change from those observed in the east. Indeed, the profiles extracted over the gas forming the C-4W cloud display CO peaks at a mean central velocity of $\sim$58~km~s$^{-1}$, all of them are correlated with \ion{H}{i} absorption dips (Fig.~\ref{COspectra}h). We interpret this result as indicative of cold \ion{H}{i} gas within a molecular structure placed at the near distance of $4.2\pm1.8$~kpc, which is absorbing the warmer \ion{H}{i} emission from the ISM at the same velocity as the CO cloud. 
Regarding the C-7.5HII zone, where the CO emission is brightest, the spectra present a peak at $\sim$56~km~s$^{-1}$ with broad emission lines of $\sim$ 9~km~s$^{-1}$ (see, for instance, Fig.~\ref{COspectra}i). 
The position of the maximum CO intensity in accordance with both the radio recombination line velocity of the \ion{H}{ii} region (58~km~s$^{-1}$, \citealt{liu2019}) and the atomic neutral gas  seen in absorption against the background provided by the \ion{H}{ii} region, led us to conclude that the molecular gas in this region forms the natal cloud of the star formation activity in G046.495$-$00.241.

Finally, we point out that our detailed analysis of the spatial distribution  of the molecular structures forming Region~C (including clouds located over a large distance interval $\sim$4-8~kpc) does not agree with the existence of a partial CO-line shell formed around {\46} at the  radial velocity of 52~km~s$^{-1}$, as proposed by \citet{sofue2021}. In their study our target source is part of an extensive list made on the basis of a morphological inspection, including around of sixty possible identification of  molecular cavities and shells towards Galactic SNRs.

\section{GeV $\gamma$ rays in the field of {\46}}
\label{gamma}
The latest version (Data Release 2) of the \it Fermi \rm Large Area Telescope Fourth Source Catalog (4FGL)\footnote{\url{https://fermi.gsfc.nasa.gov/ssc/data/access/lat/10yr_catalog/}} records an 8.1$\sigma$ detection of a $\gamma$-ray excess in the 100~MeV - 1~TeV band, labelled 4FGL~J1918.1+1215c, towards the eastern half of {\46} (first report in \citealt{abdollahi2020-fermi4thcatalog}). 
The spectrum of this source is well fit by a power law $dN/dE = \phi_0 E^{-\Gamma}$, with a differential flux $\phi_0 = (1.59 \pm 0.25) \times 10^{-12}$~cm$^{-2}$~s$^{-1}$~MeV$^{-1}$ at $E_0$ = 1099~MeV and a photon index $\Gamma = 2.65 \pm 0.08$, comparable to values measured in other SNRs with molecular cloud interactions (e.g., Kes~79, \citealt{auchettl2014}; G290.1$-$0.8, \citealt{auchettl2015}; W51C, \citealt{jogler2016}; W28, \citealt{cui2018}).

\begin{figure}[ht!]
 \includegraphics[width=0.47\textwidth]{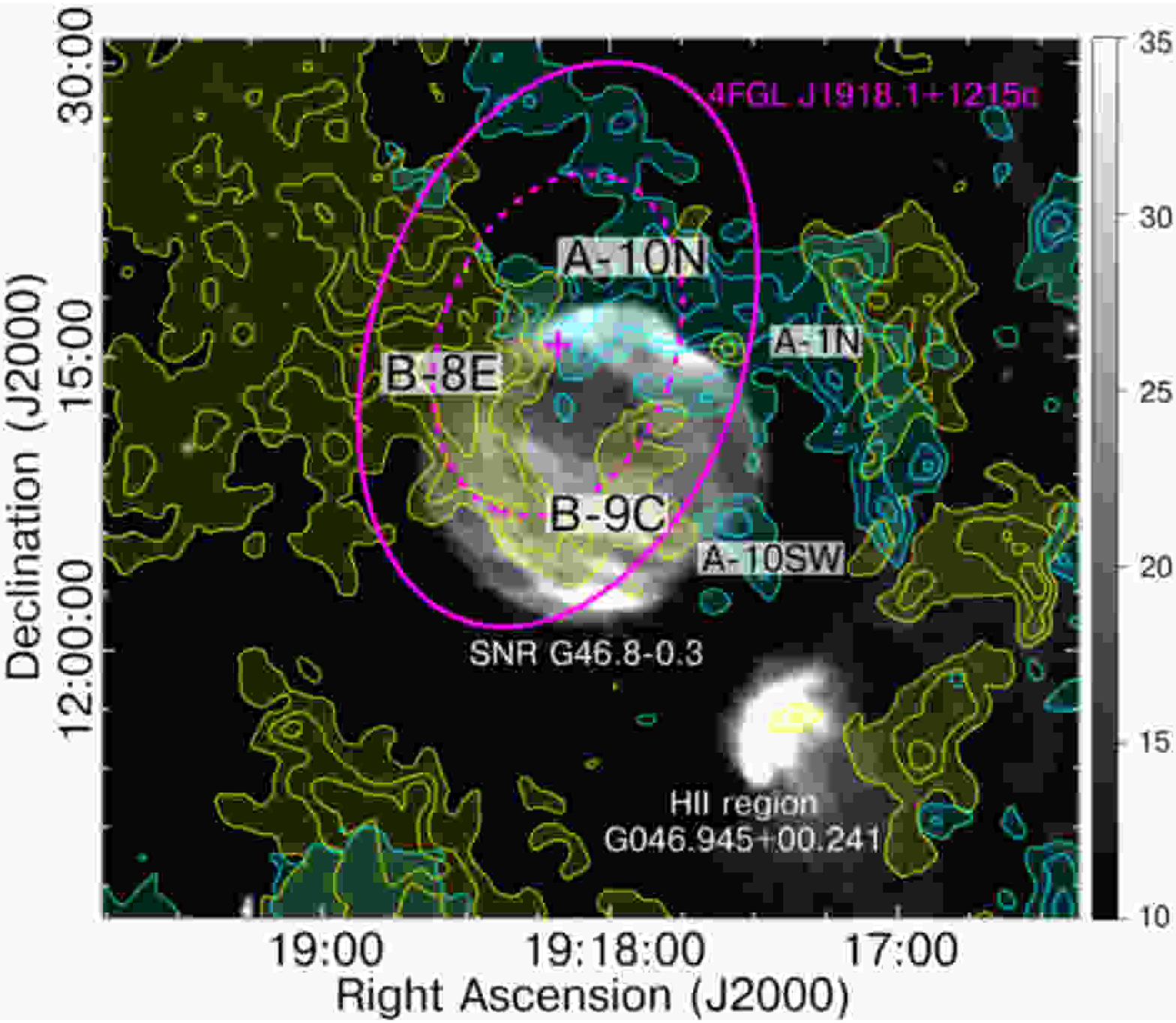}
\caption{Gray scale image (in mJy~beam$^{-1}$) corresponds to the radio continuum emission from the shell of {\46} overlaid with the \element[][12]{CO} ($J$=1-0) line emission from Regions A (in cyan, $\Delta$v=10-30~km~s$^{-1}$, levels: 40, 70, 100~K~km~s$^{-1}$) and B (in yellow, $\Delta$v=30-50~km~s$^{-1}$, levels: 55, 80, 105~K~km~s$^{-1}$) around the SNR. The velocity range of the plotted clouds is the same as in Fig.~\ref{CO-figure}. The plus symbol and ellipses illustrate the best localisation and the 95\%/68\% confidence contours of the $\gamma$-ray emission detected by \it Fermi\rm-LAT. The enclosed molecular areas corresponding to the A-10N, B-8E, and B-9C clouds  were used in Sect.~\ref{gamma} to calculate the total proton content.}
\label{gamma-elipse}
\end{figure}

Neither pulsars nor their associated wind nebulae have been detected towards the $\gamma$-ray region in {\46}, which could explain the GeV emission. Motivated by the  spatial correlation projected in the sky between the molecular cloud-SNR interaction zone and the \it Fermi \rm source, we tested the possibility that the observed $\gamma$-ray flux results from neutral pion decay emission produced by collisions of hadronic  cosmic rays with dense targets in the surrounding medium. To investigate this matter, given the limited spatial resolution of the $\gamma$-ray observations, we consider the large region $\sim30^{\prime} \times 19^{\prime}$ in size corresponding to the  95\% confidence level (CL) of the GeV source location (for comparison, the diameter of the SNR is $\sim17^{\prime}$). 
As illustrated in Fig.~\ref{gamma-elipse} the area of interest comprises portions of the molecular material embedded in the A-10N, B-8E, and B-9C ambient clouds unveiled in this work. To estimate the proton column density $N_{\mathrm{p}}$ we computed the integrated column density of the $\mathrm{H}_2$ from the $^{12}$CO line emission within the velocity ranges corresponding to these interacting clouds (see discussion in Sect.~\ref{CO}). 
Due to the doubtful association of {\46} with the molecular material forming the C-6E region, we have decided not to include any contribution from this matter in the following estimation. Using the relations presented in Sect.~\ref{ism} we obtained  $N_\mathrm{p} \simeq 2N_{\mathrm{H_{2}}} \simeq  2 \times 10^{22}$~cm$^{-2}$. 
Next, we estimated the proton density $n_{\mathrm{p}}= N_{\mathrm{p}}/L$  by adopting a path-length $L\sim 24^{\prime}.5\simeq64$~pc  
through the $\gamma$-ray emission, equivalent to the mean of the major and minor axes of the 95\% CL ellipse that we assumed is located at 9~kpc (that is, the average distance for the interacting A-10N, B-8E, and B-9C molecular clouds. See Table~\ref{molecular-table}). The resulting total proton content is then $n_{\mathrm{p}} \sim$ 100~cm$^{-3}$. 
Regarding a possible contribution from the atomic gas phase, since the analysis of the \ion{H}{i} distribution around the location of {\46} did not reveal convincing structures associated with the source (see Sect.~\ref{HI}),  it was not incorporated in the calculation above. For this reason we are aware that the derived proton density for the GeV region represents a lower limit. When taking into account the contribution (integrated in the region of the molecular clouds) of the \ion{H}{I} gas to the proton density, we find an increase by 30\%. Nevertheless, the exclusion of the atomic component does not change significantly our conclusion presented below on the energetics of cosmic rays in the region of {\46}. We also mention that our analysis does not incorporate the contribution of the ionised hydrogen to the column density since we did not find significant infrared counterparts in the region of interest.

The amount of energy injected into accelerated hadrons $W^{\mathrm{tot}}_{\mathrm{p}}$ can be estimated from the relation between the luminosity of the GeV source $L_{\gamma}$ and the characteristic cooling time of protons through the neutral mesons production channel $t_{\mathrm{pp} \rightarrow \pi^{0}}$, which in turns depends on the proton density \citep{aharonian2006}. At the adopted distance $d=9$~kpc, we obtained $L_{\gamma}(E>1~\mathrm{GeV})=4\,\pi\,d^{2}\,w_{\gamma} \simeq 5 \times 10^{34}$~erg~s$^{-1}$, where $w_{\gamma}\simeq5\times 10^{-12}$~erg~cm$^{-2}$~s$^{-1}$ is the $\gamma$-ray energy flux for $E>1$~GeV. 
On the other hand, for the total proton density we measured in the 95\% CL region of the \it Fermi \rm detection,  the cooling time of protons results to be $t_{\mathrm{pp}\rightarrow\pi^{0}}\approx 1.4 \times 10^{8} (n_{\mathrm{p}}/\mathrm{cm^{3}})^{-1}$~yr $ \sim 1.4 \times 10^{6}$~yr. Therefore, the total energy of cosmic rays required to produce the observed $\gamma$ rays flux in the region of \46~SNR is $W^{\mathrm{tot}}_{\mathrm{p}} \sim t_{\mathrm{pp}\rightarrow\pi^{0}} \, L_{\gamma} \simeq 2.2 \times 10^{48}$~erg. This value for $W^{\mathrm{tot}}_{\mathrm{p}}$ corresponds to $\sim$ 0.2\% of the canonical 
$E\simeq10^{51}$~erg SN energy and provides support to the interpretation that \46 could be the radio counterpart to the \it Fermi \rm source produced via a hadronic process. 
We notice that a similar condition remains valid for a less conservative case of analysis in a smaller region corresponding to the 68\% CL of the $\gamma$-ray detection.

Taking into account our estimation of the {\46}'s age and the properties of its environment  revealed in this work, our value for $W^{\mathrm{tot}}_{\mathrm{p}}$ accords well with those obtained for other interacting SNRs spatially correlated with $\gamma$-ray emission at GeV and/or TeV energies. We encourage the reader to consult the detailed information about this topic presented by \citet{sano2021}. It is important to note that though relatively simplistic, our analysis suggests that {\46} is out of the linear correlation proposed by these authors between the energy of cosmic rays protons and the age of the accelerator. 
Certainly, the existence of any evolutionary tendency is an important issue that should be investigated on the basis of a better statistics. Additionally, a more comprehensive study involving effects due to particle diffusion away from the SNR shell and/or re-acceleration of pre-existing cosmic rays inside the shock-compressed clouds needs to be tested to better explain the $\gamma$-ray emission from {\46}.

Finally, with the abundant interstellar matter in the vicinity of \46 still in mind, we cautiously note that on the basis of the current information on the GeV $\gamma$-ray flux in direction to the remnant a contribution to the high energy emission  from bremsstrahlung $\gamma$-rays of relativistic electrons could be also likely. 
The critical aspect in modelling the occurrence of this physical process is the requirement of high electron-proton ratios of about 0.2, larger than the 0.01 value estimated from cosmic-ray abundances \citep{gaisser1998}. Even with this caveat though, there are some examples in the literature where electron-electron or electron-ion interactions partially explain the $\gamma$ rays in the \it Fermi\rm-LAT energy band via bremsstrahlung radiation (see for instance Kes~41, \citealt{supan2018}; HB21, \citealt{ambrogi2019}). 
For the environment of the SNR~\46 the cooling timescale due to bremsstrahlung losses results $t_{\mathrm{br}} \approx 4 \times 10^{7} \, (n_{\mathrm{p}}/\mathrm{cm}^{3})^{-1}$~yr $\approx 4 \times 10^{5}$~yr \citep{aharonian2004}. 
This time is larger than the estimated age of the SNR (see Sect.~\ref{age}) and leads to a still realistically  total relativistic particle energy $W \sim 6.3 \times 10^{47}$~erg.

\section{Summary and conclusions}
\label{summary}
This work focused on determining the properties of the radio continuum emission from the SNR~\46, along with those of the atomic and molecular constituents in the surrounding medium. 
By assimilating new integrated flux densities, measured by us from recently published 88-200~MHz GLEAM maps and other ones from surveys at 1.4, 4.8, and 10~GHz, into a collection of data carefully selected from the literature, we have constructed the most complete version of the {\46}'s radio spectra available to date covering a broad range in frequency (30.9~MHz-11.2~GHz). 
A simple power-law with an $\alpha = -0.535 \pm 0.012$ slope provides the best fit to the set of fluxes measured over the remnant in the radio domain. The straightness of the spectrum implies there is no ionised gas localised either inside the remnant, in its proximity or in more distant ISM  along the line of sight able to absorb the radio emission from {\46}, which would cause a deviation from a power law at the lowest frequencies. 
Although the integrated spectrum is well-fit by a simple power-law, the analysis of the local variations in the radio spectral index with position and frequency over \46 suggests a possible steepening at approximately 1~GHz. 
To determine the reality of any potential spectral curvature within small regions over {\46} is necessary to analyse high quality radio data over a large frequency range correctly matched in the \it uv\rm-domain.

No \ion{H}{i} structures were found physically linked to {\46}. On the basis of neutral hydrogen absorption/emission spectra, we determined that the remnant is located at a distance of $8.7 \pm 1.0$~kpc, and from evolutionary models we also argued that {\46} was created in a stellar explosion likely occurred $\sim 1 \times10^4$~yr ago. 

Additionally, we provided for the first time a robust estimation for the distance to the molecular clouds lying in direction to the SNR and presented compelling signatures in the \element[][12]{CO} and \element[][13]{CO} ($J$=1-0) line emissions of interaction between the {\46}'s forward shock and dense (with $\sim100\times10^{3}$~$M_{\odot}$ and $\sim$50~cm$^{-3}$ average mass and density, respectively) molecular clouds distributed on the centre of the remnant, as well as along its northern, eastern, and southeastern edges. 
According to our findings {\46} adds to the list of evolved in our Galaxy with signs of interaction with molecular structures. Line emission observations of higher CO transitions with respect to the $J$=1-0 rotational line are of interest to quantitatively analyse variations in CO line ratios resulting from both environmental and local shock conditions in the clouds.  Unfortunately, at the time we made the analysis there were no images available from  The CO High-Resolution Survey (COHRS) $^{12}$CO ($J$ = 3–2) maps covering the complete extension of the remnant.

In line with the evidences we presented in this work on the physical relationship between {\46} and its surroundings, we also discussed and found plausible an explanation of the $\gamma$-ray flux at GeV energies via neutral pion decays after hadronic collisions, as this process requires only a few percent of the total energy released in the stellar  explosion creating {\46}. Furthermore, although very simplistic, our analysis also indicates that electron-dense medium interactions resulting in high energy bremsstrahlung radiation cannot be conclusively ruled out at this stage. 
The compilation of the radio fluxes that we have presented here is necessary as an anchor at the lowest energies for the  broad-band spectral energy distribution of \46. Its inclusion is crucial to characterise the physical process responsible for the high-energy particle production in this remnant, which we demonstrate belongs to the reduced group of evolved interacting SNRs. Future $\gamma$-ray 
observations providing complementary data with and increased sensitivity and spatial resolution in the multi-GeV \it Fermi \rm energy range and at higher energies as well, are also of interest to properly determine the spectral energy form for \46.

\begin{acknowledgements}
The authors acknowledge the anonymous referee for his/her helpful comments. G. Castelletti and L. Supan are members of the {\it Ca\-rre\-ra del Investigador Cient{\'i}fico} of CONICET, Argentina. This work was supported by the ANPCyT (Argentina) research project with number BID PICT 2017-3320. Nobeyama Radio Observatory is a branch of the National Astronomical Observatory of Japan, National Institutes of Natural Sciences. Part of the data were retrieved from the JVO portal (\url{http://jvo.nao.ac.jp/portal/}) operated by ADC/NAOJ. The National Radio Astronomy Observatory is a facility of the National Science Foundation operated under cooperative agreement by Associated Universities, Inc. 
\end{acknowledgements}

\bibliographystyle{aa}
 \bibliography{biblio-g46}
\end{document}